\documentclass[aps,prl,twocolumn,groupedaddress]{revtex4}

\usepackage{graphicx}

\begin{document}

\title{Isotropic-Nematic Transition in Liquid-Crystalline Elastomers}

\author{Jonathan V. Selinger}
\author{Hong G. Jeon}
\author{B. R. Ratna}
\affiliation{Center for Bio/Molecular Science and Engineering,
Naval Research Laboratory, Code 6900,\\
4555 Overlook Avenue, SW, Washington, DC 20375}

\date{September 24, 2002}

\begin{abstract}
In liquid-crystalline elastomers, the nematic order parameter and the induced
strain vary smoothly across the isotropic-nematic transition, without the
expected first-order discontinuity.  To investigate this smooth variation, we
measure the strain as a function of temperature over a range of applied stress,
for elastomers crosslinked in the nematic and isotropic phases, and analyze the
results using a variation on Landau theory.  This analysis shows that the
smooth variation arises from quenched disorder in the elastomer, combined with
the effects of applied stress and internal stress.
\end{abstract}


\maketitle

Liquid-crystalline elastomers are unusual materials that combine the elastic
properties of rubbers with the anisotropy of liquid
crystals~\cite{warner96,terentjev99}.  They consist of crosslinked networks of
polymers with mesogenic units.  Because of this structure, any stress on the
polymer network influences the orientational order of the liquid crystal, and
conversely, any change in the orientational order affects the shape of the
elastomer.  These materials are being actively studied for both basic
research~\cite{mao00,warner00,clarke01prl,mao01,finkelmann01} and applications,
including use as actuators or artificial muscles~\cite{thomsen01}.  For this
application, a change in temperature near the isotropic-nematic transition
induces a large change in the orientational order, which causes the elastomer to
extend or contract.

In this paper, we investigate the isotropic-nematic transition in
liquid-crystalline elastomers.  In conventional liquid crystals, this is a
first-order transition, with a discontinuity in the magnitude of the
orientational order as a function of temperature.  By contrast, experiments on
liquid-crystalline elastomers show that both the orientational order parameter
and the elastomer strain change smoothly at this transition, with no first-order
discontinuity~\cite{thomsen01,schatzle89,kaufhold91,disch94,clarke01pre}.
Surprisingly, this is neither a first- nor a second-order transition, but rather
a rapid nonsingular crossover from the isotropic to the nematic phase.  Thus,
the key question is how to explain this difference between conventional liquid
crystals and liquid-crystalline elastomers.  That question is important for
basic research, because it shows how orientational ordering is affected by
coupling to a crosslinked polymer network.  That question is also important for
applications, because it shows how to optimize these materials for artificial
muscles, which should have the greatest possible length change for a fixed
temperature change.

\begin{figure}[b]
\includegraphics[clip,width=2.74in]{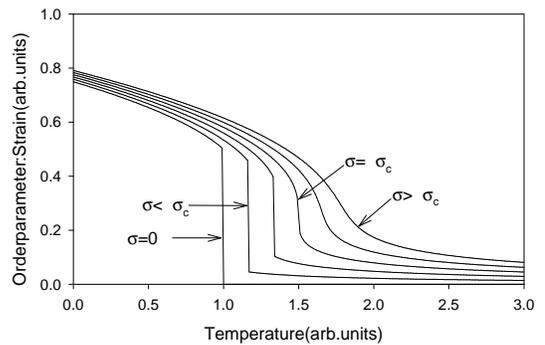}%
\caption{\label{generic}Prediction for the strain as a function of temperature
in a \emph{homogeneous} elastomer under an aligning stress $\sigma$.}
\end{figure}

There are two possible explanations for a smooth crossover from the isotropic to
the nematic phase.  The first explanation is based on the effect of an aligning
stress on a first-order transition~\cite{degennes75}.  The classical theory of
phase transitions predicts the generic behavior shown in Fig.~\ref{generic}.
For a stress below the critical point, a system has a first-order discontinuity
in the order parameter at the transition.  As the stress increases, the
discontinuity decreases.  When the stress reaches the critical point, the
discontinuity vanishes and the system has an infinite slope in the order
parameter as a function of temperature.  Beyond the critical point, the system
evolves smoothly from the disordered to the ordered phase.  In
liquid-crystalline elastomers, an aligning stress may come from an applied
stress on the sample.  It may also come from an internal stress due to
crosslinking an elastomer in the nematic phase, which imprints orientational
order in the pattern of crosslinks.  It is possible that the combination of
applied stress and internal stress might put an elastomer beyond the critical
point, so that it would show a supercritical evolution from the isotropic to the
nematic phase.  If this explanation is correct, then one would optimize
elastomers for applications by working close to the critical point, where the
slope is greatest.

An alternative explanation for this behavior is heterogeneity in an elastomer.
The polymerization and crosslinking process induces some quenched disorder in a
sample.  For example, polydispersity in the chain length gives one type of
disorder.  This disorder may lead to a distribution of regions with different
isotropic-nematic transition temperatures.  In that case, at any given
temperature, a sample would have a coexistence of isotropic and nematic domains.
As the temperature decreases, it would cross over from mostly isotropic to
mostly nematic, leading to a smooth evolution in the average orientational order
parameter and in the macroscopic strain.  If this explanation is correct, then
one would optimize elastomers for applications by reducing the heterogeneity to
get the transition in the narrowest possible range of temperature.

\begin{figure}
\includegraphics[clip,width=2.74in]{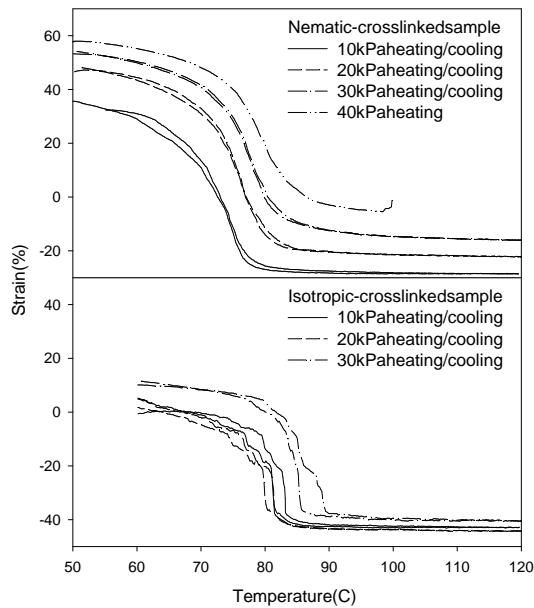}%
\caption{\label{data}Data for strain as a function of temperature over a range
of applied stress, for elastomers crosslinked in the nematic and isotropic
phases.  Strain is measured relative to an arbitrary zero value.  The
nematic-crosslinked sample broke during the 40 kPa heating run.}
\end{figure}

To determine which explanation is correct, we measure the strain as a function
of temperature over a range of applied tensile stress.  We use elastomer samples
crosslinked in the nematic phase, which should have a large internal stress
imprinted by the crosslinking process, and samples crosslinked in the isotropic
phase, which should not have an internal stress.  The samples are composed of a
50/50 mol \% mixture of 4$^\prime$-acryloyl\-oxy\-butyl 2,5-(4$^\prime$-butyl\-%
oxy\-benzoyl\-oxy)benzoate (MAOC4) and 4$^\prime$-acryloyl\-oxy\-butyl 2,5-%
di(4$^\prime$-pentyl\-cyclo\-hexyloyl\-oxy)benzoate (MACC5), with 5 mol \% of
the 1,6-hexanediol diacrylate crosslinker.  Nematic-crosslinked
samples are simultaneously polymerized and crosslinked in a cell with rubbed
surfaces at $30^\circ$C, and isotropic-crosslinked samples at $110^\circ$C.
Synthesis, preparation, and characterization of nematic-crosslinked samples are
described in Ref.~\cite{thomsen01}.  At low temperature those samples have
long-range orientational order, with order parameter
$S={\langle\frac{3}{2}\cos^2\theta-\frac{1}{2}\rangle}$ of 0.3, as determined by
polarized FTIR spectroscopy~\cite{thomsen01}.

Under zero applied stress, the samples were seen to extend and contract as a
function of temperature.  For finite applied stress, the thermoelastic curves
for strain vs.\ temperature were obtained in static
measurements on a dynamic mechanical analyzer (TA Instruments DMA 2980) at a
heating/cooling rate of $0.5^\circ$C/min.  The data are shown in
Fig.~\ref{data}.  The strain is measured relative to an arbitrary zero value.
These plots show a smooth nonsingular isotropic-nematic transition at all values
of the applied stress, and under both crosslinking conditions, although the
transition is sharper at lower applied stress and under isotropic crosslinking.

To assess whether the data are compatible with the first proposed explanation,
we use Landau theory for a homogeneous elastomer.  The free energy can
be expanded in terms of the orientational order parameter
$S$ and the strain $e$ relative to the high-temperature relaxed state.  This
expansion gives~\cite{degennes75}
\begin{equation}
F = \textstyle\frac{1}{2}\alpha'(T-T'_0)S^2 - \textstyle\frac{1}{3}b'S^3
+ \textstyle\frac{1}{4}c'S^4 - u e S - \sigma e + \textstyle\frac{1}{2}\mu e^2 ,
\end{equation}
where $T$ is the temperature and $\sigma$ is the effective stress acting on the
elastomer, which is a combination of the applied stress and the internal stress
due to anisotropic crosslinking.  We average this free energy over $S$ to obtain
the free energy in terms of $e$ alone,
\begin{equation}
F = \textstyle\frac{1}{2}\alpha(T-T_0)e^2 - \textstyle\frac{1}{3}b e^3
+ \textstyle\frac{1}{4}c e^4 - \sigma e .
\label{effectivefreeenergy}
\end{equation}
Minimizing this free energy over $e$ gives
\begin{equation}
\alpha(T-T_0)e - b e^2 + c e^3 - \sigma = 0 .
\label{implicit}
\end{equation}
This equation implicitly determines $e$ as a function of $T$ and $\sigma$.  In
particular, it predicts a critical point at
\begin{equation}
\sigma_c = \frac{b^3}{27c^2},\ T_c = T_0 + \frac{b^2}{3c\alpha}.
\label{criticalpoint}
\end{equation}

We attempt to fit the prediction of Eq.~(\ref{implicit}) to the data presented
above.  For computational convenience, we solve Eq.~(\ref{implicit}) for the
inverse function $T(e)$ and fit it to temperature as a function of strain.
Because this inverse function depends linearly on the fitting parameters, we can
use linear regression techniques.  Since the strain data are reported relative
to an arbitrary zero, we subtract off the high-temperature asymptotic strain
from the data to obtain the values of $e$ for the analysis.  This procedure fits
the data to the functional form shown in Fig.~\ref{generic}.  The fit can be on
either the first-order or the supercritical side of the critical point,
depending on the data.

\begin{figure}
\includegraphics[clip,width=2.74in]{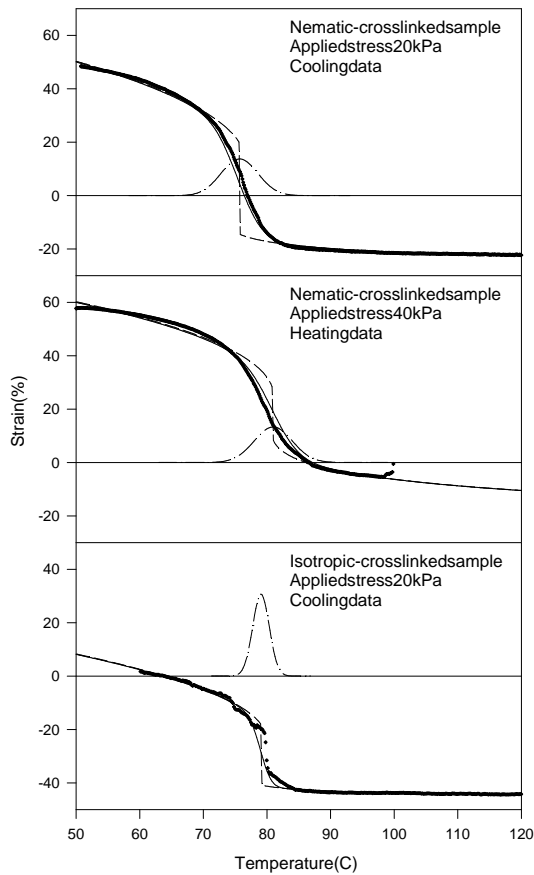}%
\caption{\label{fits}Fits of selected data sets to the models discussed in the
text.  Dashed lines:  Homogeneous model.  Solid lines:  Heterogeneous model.
Dot-dashed lines:  Distribution of the transition temperature $T_{NI}$ in the
heterogeneous model.}
\end{figure}

Our analysis gives the fits shown by the dashed lines in Fig.~\ref{fits}.  These
fits agree well with the data in the high- and low-temperature limits, but they
are unsatisfactory for intermediate temperatures.  In all cases, the fitting
function shows a first-order discontinuity at the isotropic-nematic transition.
Apparently the regression sacrifices the intermediate regime in order to give
good fits at high and low temperatures.  We also fit the data to an extended
model with fifth- and sixth-order terms in the free energy (not shown), and the
fits are also unsatisfactory.  These results show that the data are not
consistent with a supercritical evolution between the isotropic and nematic
phases.  The high- and low-temperature data do not connect together following
the prediction for a supercritical transition in a homogeneous elastomer.

In addition to the unsatisfactory fits, there are two other indications that the
data are inconsistent with predictions for a supercritical transition in a
homogeneous elastomer.  First, we can use the Landau theory to extract the
maximum slope $\partial e/\partial T$ at the inflection point in the
supercritical regime $\sigma>\sigma_c$.  The result is
\begin{equation}
\left(\frac{\partial e}{\partial T}\right)_\mathrm{max} =
- \frac{\alpha}{3c^{2/3}(\sigma^{1/3}-\sigma_c^{1/3})} .
\end{equation}
This equation implies that the maximum slope should decrease inversely with
stress $\sigma-\sigma_c$ beyond the critical point.  However, the measured slope
depends rather weakly on stress, and it is approximately constant with respect
to stress at low stress.  This result suggests that some mechanism other than
supercritical stress is responsible for the observed broadening of the
transition.  Second, in this experiment we minimize any symmetry-breaking
influence on the elastomer by crosslinking one sample in the isotropic phase and
reducing the applied stress to 10 kPa.  Even under these conditions, the
experiment shows a smooth crossover between the isotropic and nematic phases.
It is unlikely that these conditions could give a supercritical stress on the
system; it is more plausible that another mechanism is involved.

Because of these inconsistencies between the data and the model for a
homogeneous elastomer, we consider a model for heterogeneity in the elastomer.
As a hypothesis, we suppose that heterogeneity gives regions with different
isotropic-nematic transition temperatures.  We consider a Gaussian distribution
of the transition temperature $T_{NI}$, and thus of the parameter $T_0$.  Hence,
the macroscopic strain is an average over the strain of local regions, which can
be written as the convolution
\begin{equation}
e_\mathrm{hetero}(\sigma,T)= \int d(T_0) e_\mathrm{homo}(\sigma,T-T_0) P(T_0),
\end{equation}
where
\begin{equation}
P(T_0)=\frac{1}{\sqrt{2\pi}T_{0,SD}}
\exp\left[-\frac{1}{2}\left(\frac{T_0-\overline{T_0}}{T_{0,SD}}\right)^2\right].
\end{equation}

To compare this heterogeneous prediction with the data, we take the homogeneous
fit discussed above and convolve it with a Gaussian of adjustable width
$T_{0,SD}$.  We fit the width to the data through a nonlinear least-squares
procedure.  The results are shown by the solid lines in Fig.~\ref{fits}.  The
convolution does not change the high- or low-temperature limits of the fits,
which were already satisfactory, but it has a great effect on the
intermediate-temperature behavior.  Instead of a discontinuous jump in the
strain at $T_{NI}$, the fitted curves show a smooth crossover as the elastomer
changes from mostly isotropic to mostly nematic.  The shape of the curve in the
intermediate-temperature regime is approximately an error function of width
$T_{0,SD}$.  This behavior agrees with the trend in the data.  As a result, the
heterogeneous model gives good fits over the full range of temperature.

\begin{figure}
\includegraphics[clip,width=2.74in]{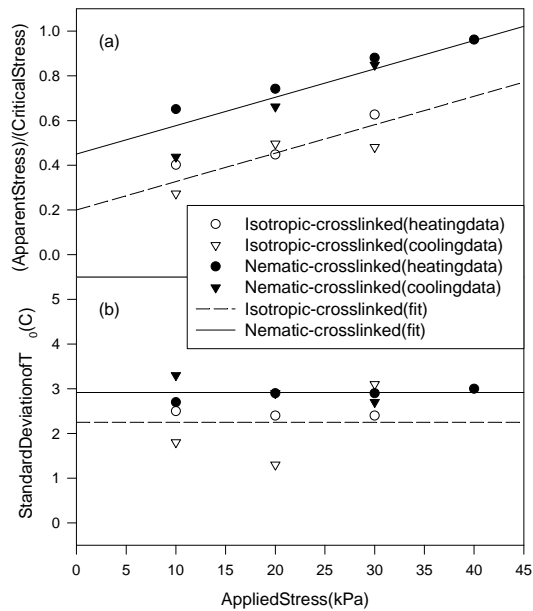}%
\caption{\label{analysis}Analysis of the fitting results for isotropic- and
nematic-crosslinked elastomers over a range of applied stress.  (a) Stress ratio
$\sigma/\sigma_c$, indicating how close an elastomer is to the critical point.
(b) Standard deviation $T_{0,SD}$, indicating the width of the distribution of
transition temperatures.}
\end{figure}

To analyze the fitting results further, we would like to know how close an
elastomer is to the critical point at $\sigma=\sigma_c$.  For that reason, we
define the dimensionless stress ratio $\sigma/\sigma_c$, where $\sigma$ is the
apparent stress that comes from fitting the homogeneous model to the high- and
low-temperature data, and $\sigma_c=b^3/(27c^2)$ is the critical stress derived
from the parameters $b$ and $c$ in those fits.  In Fig.~4(a) we plot the stress
ratio for isotropic- and nematic-crosslinked samples vs.\ applied stress.  From
this plot, we can make several observations.  First, the stress ratio increases
linearly with applied stress.  This increase is expected, since an applied
stress should shift a sample toward the critical point.  At the highest applied
stress of 40 kPa, the nematic-crosslinked sample nearly reaches the critical
point.  When the applied stress goes to zero, the stress ratio does not go to
zero but rather to a finite limit.  This behavior shows that the effective
stress acting on a sample is a combination of the applied stress and an internal
stress.  For the nematic-crosslinked sample, we expect an internal stress due to
the anisotropy of the crosslinking, and indeed the internal stress is large,
giving a contribution equivalent to 35 kPa of applied stress.  For the
isotropic-crosslinked sample, we do not expect an internal stress, but Fig.~4(a)
implies that some internal stress is present, equivalent to 16 kPa of applied
stress.  This surprising contribution must come from some unintentional
anisotropy in the sample preparation, perhaps related to the boundary conditions
on the sample.

Another parameter to extract from the fits is the Gaussian width $T_{0,SD}$
that is required to fit the isotropic-nematic crossover in the data.  In
Fig.~4(b) we plot $T_{0,SD}$ vs.\ applied stress for isotropic- and
nematic-crosslinked samples.  The plot shows that the fitted values of
$T_{0,SD}$ have some scatter, but they are not correlated with applied stress.
This result is reasonable, because the distribution of transition temperatures
should not be related to applied stress.  The average value of $T_{0,SD}$ is
approximately $2.25^\circ$C for the isotropic- and $2.9^\circ$C for the
nematic-crosslinked sample.  This corresponds to a distribution with a full
width at half maximum of $5.3^\circ$C or $6.8^\circ$C in each of the samples,
respectively.

As a final point, we note that our model for a distribution of transition
temperatures describes only one way in which heterogeneity can affect
liquid-crystalline elastomers.  A second possible mechanism would be a
distribution in the direction of the imprinted orientational order, especially
in a sample crosslinked in the nematic phase.  Indeed, a distribution of
quenched director orientations might explain the greater value of the Gaussian
width found in the nematic- than in the isotropic-crosslinked sample.  A
distribution of transition temperatures is random-bond disorder, while a
distribution of quenched director orientations is random-field disorder.  Our
study has shown that the sharpness of the isotropic-nematic transition is
controlled by heterogeneity, but it has not addressed the question of whether
random-bond or random-field disorder is dominant.  Indeed, measurements of the
strain vs.\ temperature may not be enough to make this distinction; more
microscopic studies may be needed.  This remains a question for future research.

In conclusion, we have developed a phenomenological theory for the
isotropic-nematic transition in liquid-crystalline elastomers.  This theory is a
variation on Landau theory, which allows for quenched disorder in the elastomer
through variation in the transition temperature.  We compare this theory with
measurements of the strain vs.\ temperature over a range of applied stress, for
samples crosslinked in the isotropic and nematic phases.  This comparison shows
that applied stress, internal stress, and quenched disorder are \emph{all}
involved in the shape of the thermoelastic curve.  In particular, quenched
disorder is a key limiting factor in the sharpness of the isotropic-nematic
transition, which must be controlled for applications of liquid-crystalline
elastomers.

We thank P. Keller for synthesizing the materials.  This work was supported by
the Office of Naval Research and the Defense Advanced Research Projects Agency.
HGJ was supported by the National Research Council Associateship Program.


\begin{thebibliography}{}

\bibitem{warner96} M.~Warner and E.~M. Terentjev, Prog. Polym. Sci. \textbf{21},
853 (1996).

\bibitem{terentjev99} E.~M. Terentjev, J. Phys.-Condens. Matter \textbf{11},
R239 (1999).

\bibitem{mao00} Y. Mao and M. Warner, Phys. Rev. Lett. \textbf{84}, 5335 (2000).

\bibitem{warner00} M. Warner, E.~M. Terentjev, R.~B. Meyer, and Y. Mao, Phys.
Rev. Lett. \textbf{85}, 2320 (2000).

\bibitem{clarke01prl} S.~M. Clarke, A.~R. Tajbakhsh, E.~M. Terentjev, and M.
Warner, Phys. Rev. Lett. \textbf{86}, 4044 (2001).

\bibitem{mao01} Y. Mao and M. Warner, Phys. Rev. Lett. \textbf{86}, 5309 (2001).

\bibitem{finkelmann01} H. Finkelmann, E. Nishikawa, G.~G. Pereira, and M.
Warner, Phys. Rev. Lett. \textbf{87}, 015501 (2001).

\bibitem{thomsen01} D.~L. Thomsen \textit{et al.}, Macromolecules \textbf{34},
5868 (2001).

\bibitem{schatzle89} J.~Sch\"atzle, W.~Kaufhold, and H.~Finkelmann, Makromol.
Chem. \textbf{190}, 3269 (1989).

\bibitem{kaufhold91} W.~Kaufhold, H.~Finkelmann, and H.~R. Brand, Makromol.
Chem. \textbf{192}, 2555 (1991).

\bibitem{disch94} S.~Disch, C.~Schmidt, and H.~Finkelmann, Macromol. Rapid
Commun. \textbf{15}, 303 (1994).

\bibitem{clarke01pre} For an exception, see S.~M. Clarke, A. Hotta, A.~R.
Tajbakhsh, and E.~M. Terentjev, Phys. Rev. E \textbf{64}, 061702 (2001).  In
this case, the strain seems to increase as a power law below a second-order
isotropic-nematic transition.  That behavior has not yet been explained
theoretically, and we do not address it here.

\bibitem{degennes75} P.~G. de~Gennes, C. R. Acad. Sci. Ser. B \textbf{281}, 101
(1975).

\end{thebibliography}
\end{document}